\documentstyle[prd,aps,preprint,tighten,epsfig]{revtex}

\newcommand{\ccontra}[2]{\begin{picture}(38,10)
  \put(0,0){$#1$}
  \put(2,-7){\line(0,1){4}}
  \put(2,-7){\line(1,0){35}}
  \put(37,-7){\line(0,1){4}}
  \put(40,-9){\tiny B(t)}
  \put(35,0){$#2$} 
\end{picture}}

\newcommand{\ccontrad}[2]{\begin{picture}(44,10)
  \put(0,0){$#1$}
  \put(2,-7){\line(0,1){4}}
  \put(2,-7){\line(1,0){41}}
  \put(43,-7){\line(0,1){4}}
  \put(45,-9){\tiny B(t)}
  \put(41,0){$#2$} 
\end{picture}}

\newcommand{\ccontradd}[2]{\begin{picture}(47,10)
  \put(0,0){$#1$}
  \put(2,-7){\line(0,1){4}}
  \put(2,-7){\line(1,0){44}}
  \put(46,-7){\line(0,1){4}}
  \put(48,-9){\tiny B(t)}
  \put(44,0){$#2$} 
\end{picture}}

\begin{document}
\draft
\title{Derivation of Transport Equations using the Time-Dependent Projection
Operator Method}
\author{Tomoi Koide\thanks{tkoide@yukawa.kyoto-u.ac.jp}}
\address{Yukawa Institute for Theoretical Physics, Kyoto University,
Kyoto 606-8502, Japan}
\date{Received November 8, 2001} \maketitle

\begin{abstract}
We develop a formalism to carry out coarse-grainings in 
quantum field theoretical systems 
by using a time-dependent projection operator in the Heisenberg picture.
A systematic perturbative expansion with respect to the interaction
part of the Hamiltonian is given, 
and a Langevin-type equation without a time-convolution 
integral term is obtained.
This method is applied to a quantum field theoretical model, 
and coupled transport equations are derived.
\end{abstract}

\section{Introduction}

Obtaining descriptions of nonequilibrium processes in quantum field theory 
is an important problem and has been actively studied.
In the treatment of such systems, 
it is often important to carry out reductions or coarse-grainings 
of irrelevant degrees of freedom.
In classical systems, 
it is widely believed that a system of macroscopic size composed of 
many microscopic variables exhibits rather simple macroscopic behavior 
that can be 
described in terms of only a few macroscopic variables.
It is therefore not unreasonable to conjecture 
that such an elimination of the irrelevant 
information can be developed for application to quantum systems 
and help in the analysis of nonequilibrium quantum processes.
The projection operator method is one well-known method 
for carrying out coarse-grainings systematically.
\cite{ref:NZ,ref:Mori,ref:Toku-Mori,ref:HSS,ref:Koi-Mar}
In this method, after the elimination of irrelevant information 
by means of a projection operator, 
a kind of master equation in the Schr\"odinger picture 
or a Langevin-type equation in the Heisenberg picture is obtained.
Famous examples are the Nakajima-Zwanzig and Mori equations.
\cite{ref:NZ,ref:Mori}
Recently, the unification and generalization of the treatments 
used in these two picture has been realized.\cite{ref:HSS,ref:Koi-Mar}

In the derivation of kinetic equations from the microscopic point of view, 
we usually assume that the relevant part of the system 
interacts with an irrelevant subsystem, which is regarded as 
a heat reservoir in thermal equilibrium.\cite{ref:lan}
However, in a general nonequilibrium system, we can expect 
the irrelevant subsystem to have some time dependence.
Furthermore, it is often convenient to have a description 
in which two or more coupled systems are considered to 
be on an equal footing.
In particular, in quantum field theory, 
both relevant and irrelevant parts of the system may be composed of 
infinite numbers of degrees of freedom, and 
therefore, it is not clear whether we can regard one subsystem as a reservoir 
for the other or not.
To elucidate this situation, it is convenient to introduce 
a time-dependent projection operator.

The time-dependent projection operator was first applied by Robertson.
\cite{ref:Rob}
He attempted to project the complex behavior of the relevant density 
matrix onto a local equilibrium density matrix 
and derived a master equation.
Ochiai attempted to obtain a kinetic equation 
by using a similar projection.\cite{ref:Ochi}
Robertson's time-dependent projection operator was improved by Kawasaki et al.
\cite{ref:Kawa-Gun}
Willis et al. considered coupled systems and 
introduced another time-dependent projection operator.\cite{ref:Will-Pic}
A similar projection was implemented by Grabert et al.
\cite{ref:Gra-Wei}
Shibata et al. derived a systematic perturbative expansion formula 
for a master equation without time-convolution integral terms.
\cite{ref:Shi-Has-tp}

In all of the above cited works, formulations were developed 
in the Schr\"odinger picture, 
and master equations were derived.
Langevin-type equations are obtained 
when we apply the projection operator method in the Heisenberg picture. 
Such equations were studied by Grabert\cite{ref:Gra} 
and Furukawa.\cite{ref:Furu}
They carried out an exact transformation of the Heisenberg equation of 
motion to derive a Langevin-type equation.
However, to apply this approach to concrete phenomena, 
such an exact formulation 
is not convenient, and we must resort to perturbative calculations.
In this paper, we develop a systematic perturbative expansion formula 
of a Langevin-type equation.
It is known that both equations with and without 
a time-convolution integral term 
can be derived using the projection operator method.
\cite{ref:Mori,ref:Toku-Mori,ref:HSS,ref:Koi-Mar}
In this study, we investigate an equation without 
a time-convolution integral term.
Furthermore, to investigate the validity of our formalism, 
we apply it to a quantum field theoretical model and 
attempt to derive coupled transport equations.

This paper is organized as follows.
In \S \ref{sec:2}, 
the projection operator method with a time-dependent 
projection operator is developed.
In \S \ref{sec:3}, 
we apply our formalism to a quantum field theoretical model 
that is composed of two bosons.
We assume local equilibrium and derive coupled 
transport equations.
A summary and conclusions are given in \S \ref{sec:4}.

\section{Time-dependent projection operator method}\label{sec:2}

Our starting point is the Heisenberg equation of motion,
\begin{eqnarray}
  \frac{d}{dt}O(t) &=& i[H,O(t)] \\
                   &=& iLO(t) \\
   \longrightarrow O(t) &=& e^{iL(t-t_{0})}O(t_{0}), \label{eqn:HE}
\end{eqnarray}
where $L$ is the Liouville operator and $t_{0}$ is the time at which 
we prepare an initial state.
The Heisenberg equation contains complete information of the time-evolution 
of the operator, but in general, 
it is difficult to solve exactly when there are interactions.
For this reason, 
it is necessary to carry out a reduction or a coarse-graining 
of the irrelevant information.
For this purpose, we introduce the time-dependent 
projection operators $P(t)$ and $Q(t)$, which are related as 
\begin{eqnarray}
Q(t) = 1-P(t).
\end{eqnarray}
The projection operator $P(t)$ helps us to project any operator onto 
the P-space, which consists of the relevant degrees of freedom.
From Eq.~(\ref{eqn:HE}), one can see that the time dependence 
of the operators is determined by $e^{iL(t-t_{0})}$.
This yields 
\begin{eqnarray}
   \frac{d}{dt}e^{iL(t-t_{0})} &=& e^{iL(t-t_{0})}iL \nonumber \\
                &=& e^{iL(t-t_{0})}(P(t)+Q(t))iL.   \label{eqn:P+Q}
\end{eqnarray}
From this equation, we can derive the two equations
\begin{eqnarray}
\frac{d}{dt}e^{iL(t-t_{0})}P(t)
&=& e^{iL(t-t_{0})}(P(t)+Q(t))iLP(t)+e^{iL(t-t_{0})} \dot{P}(t),
\label{eqn:P} \\
\frac{d}{dt}e^{iL(t-t_{0})} Q(t)
&=& e^{iL(t-t_{0})} (P(t)+Q(t))iLQ(t)+e^{iL(t-t_{0})} \dot{Q}(t)\label{eqn:Q},
\end{eqnarray}
where $\dot{P}(t) = dP(t)/dt$ and $\dot{Q}(t) = dQ(t)/dt$.
Equation (\ref{eqn:Q}) can be solved for $e^{iL(t-t_{0})}Q$:
\begin{eqnarray}
e^{iL(t-t_{0})}Q(t)
&=& Q(t_{0})e^{i\int^{t}_{t_{0}}ds LQ(s)}_{\rightarrow}
+\int^{t}_{t_{0}}ds e^{iL(s-t_{0})}
(\dot{Q}(s)+P(s)iLQ(s))e^{i\int^{t}_{s}d\tau LQ(\tau)}_{\rightarrow}
\nonumber \\
\label{eqn:Int-Q-TC}\\
&=& Q(t_{0})e^{i\int^{t}_{t_{0}}ds LQ(s)}_{\rightarrow}
+e^{iL(t-t_{0})}
(P(t)+Q(t)) \Sigma(t,t_{0}) \nonumber \\
&=& 
\{
Q(t_{0})e^{i\int^{t}_{t_{0}}ds LQ(s)}_{\rightarrow}
+e^{iL(t-t_{0})}P(t)\Sigma(t,t_{0})  
\}\frac{1}{1-\Sigma(t,t_{0})},\label{eqn:Int-Q}
\end{eqnarray}
where  
\begin{eqnarray}
\Sigma(t,t_{0}) &=& 
\int^{t}_{t_{0}}ds e^{-iL(t-s)} 
\{
\dot{Q}(s)+P(s)iLQ(s)
\} e^{i\int^{t}_{s}d\tau LQ(\tau)}_{\rightarrow}.\label{eqn:sigma}
\end{eqnarray}
Here, the time ordered operator $e^{i\int^{t}_{t_{0}}ds LQ(s)}_{\rightarrow}$ 
is defined as
\begin{eqnarray}
e^{i\int^{t}_{t_{0}}ds LQ(s)}_{\rightarrow} &=& 
1 + \sum^{\infty}_{n=1}
i^n \int^{t}_{t_{0}}dt_{1}\int^{t_{1}}_{t_{0}}dt_{2}
\cdots\int^{t_{n-1}}_{t_{0}}dt_{n}LQ(t_{n})LQ(t_{n-1})\cdots
LQ(t_{1}). \nonumber \\
\end{eqnarray}
Note that there is a term including $\dot{Q}(s)$ 
that is not observed in the 
the time-independent projection operator method.\cite{ref:Koi-Mar}

Substituting Eq.~(\ref{eqn:Int-Q}) into Eq.~(\ref{eqn:P+Q}) 
and operating with $O(t_{0})$ from the right, 
we obtain 
\begin{eqnarray}
\frac{d}{dt} O(t) &=& e^{iL(t-t_{0})} P(t)iLO(t_{0}) \nonumber \\
&& + e^{iL(t-t_{0})}  
P(t)\Sigma(t,t_{0})\frac{1}{1-\Sigma(t,t_{0})}iLO(t_{0}) \nonumber \\
&&+Q(t_{0})e^{i\int^{t}_{t_{0}}dsLQ(s)}_{\rightarrow} 
\frac{1}{1-\Sigma(t,t_{0})}iL(t,t_{0})O(t_{0}). \label{eqn:TCL}
\end{eqnarray}
This equation has no time-convolution integral.
This point is demonstrated at the end of this section.
We call this the ``time-convolutionless'' (TCL) equation.
When the time-dependence of the projection operator is ignored, 
this equation agrees with Eq.~(2$\cdot$13) of Ref.~\cite{ref:Koi-Mar}.
When we substitute Eq.~(\ref{eqn:Int-Q-TC}) into Eq.~(\ref{eqn:P+Q}), 
we obtain an equation with a time-convolution integral term, 
which is called the ``time-convolution'' (TC) equation.
\cite{ref:HSS,ref:Koi-Mar}
However, we discuss only the TCL equation in this paper.

The TCL equation is exactly equivalent to the 
Heisenberg equation of motion.
Usually, we carry out the perturbative expansion of the interaction 
Hamiltonian and retain only the lowest order terms, which are often sufficient
to describe the dissipation effect.
However, the above TCL equation is not convenient to 
carry out the perturbative expansion.
Our next task is therefore to rewrite the TCL equation.
For this purpose, 
we restrict the nature of the projection operator.
For the time-independent projection operator $P$, the condition $P^2=P$ 
is satisfied.
Contrastingly, in the time-dependent case, the order of 
operation of the projection operators is important.
In most previous works, 
the condition $P(t_{1})P(t_{2}) = P(t_{1})$ was employed.
\cite{ref:Rob,ref:Ochi,ref:Kawa-Gun,ref:Will-Pic,ref:Gra-Wei,ref:Shi-Has-tp}\cite{ref:Furu}
However, in this paper, we assume the condition
\begin{eqnarray}
P(t_{1})P(t_{2}) = P(t_{2}), \label{eqn:PPP}
\end{eqnarray}
because the projection operator that we use in the next section 
satisfies this condition.
From this condition, we can derive several relations:
\begin{eqnarray}
Q(t_{1})P(t_{2}) &=& 0, \\
P(t_{1})Q(t_{2}) &=& P(t_{1}) - P(t_{2}), \\
Q(t_{1})Q(t_{2}) &=& Q(t_{1}). \label{eqn:QQQ}
\end{eqnarray}

The total Hamiltonian of the system can be divided into two parts,
\begin{eqnarray}
H = H_{0}(t,t_{0}) + H_{I}(t,t_{0}),
\end{eqnarray}
where $H_{0}(t,t_{0})$ and $H_{I}(t,t_{0})$ are 
the nonperturbative part and the perturbative part of the Hamiltonian, 
respectively.
In a general nonequilibrium process, 
we can consider the case in which the mass of a particle changes with time, 
and therefore, $H_{0}(t,t_{0})$ becomes time dependent.
The corresponding Liouville operators are defined as 
\begin{eqnarray}
L_{0}(t,t_{0})O = [H_{0}(t,t_{0}),O],~~~
L_{I}(t,t_{0})O = [H_{I}(t,t_{0}),O].
\end{eqnarray}
Now, we assume another condition,
\begin{eqnarray}
Q(t_{2})L_{0}(t_{1},t_{0})P(t_{1}) = 0.\label{eqn:QL0P}
\end{eqnarray}
With this condition, 
the form of the nonperturbative Hamiltonian affects 
that of the projection operator, and vice versa.
This condition means that the nonperturbative Hamiltonian 
does not drive the operator, once it is projected onto the P-space, 
into the Q-space, which is orthogonal to the P-space.

With the above properties of the projection operators, 
the function $\Sigma(t,t_{0})$ can be expressed as 
\begin{eqnarray}
\Sigma(t,t_{0})
&=& 
\int^{t}_{t_{0}}ds e^{-iL(t-s)} 
\{
\dot{Q}(s)+P(s)iLQ(s)
\} e^{i\int^{t}_{s}d\tau LQ(\tau)}_{\rightarrow} \nonumber \\
&=& Q(t) - e^{-iL(t-t_{0})}
Q(t_{0})e^{i\int^{t}_{t_{0}}d\tau LQ(\tau)}_{\rightarrow} \nonumber \\
&=& Q(t) - U_{0}^{-1}(t,t_{0}){\cal C}(t,t_{0})Q(t_{0})
{\cal D}(t,t_{0})e^{i\int^{t}_{t_{0}}ds L_{0}(s,t_{0})Q(s)}_{\rightarrow}.
\end{eqnarray} 
Here, the operators ${\cal C}(t,t_{0})$ and ${\cal D}(t,t_{0})$ 
are expressed as
\begin{eqnarray}
{\cal C}(t,t_{0})&=&U_{0}(t,t_{0})e^{-iL(t-t_{0})} \nonumber \\
&=& 1+\sum^{\infty}_{n=1}(-i)^n 
\int^{t}_{t_{0}}dt_{1}\int^{t_{1}}_{t_{0}}dt_{2}
\cdots\int^{t_{n-1}}_{t_{0}}dt_{n}
\breve{L}_{I}(t_{1},t_{0})\breve{L}_{I}(t_{2},t_{0})\cdots\breve{L}_{I}(t_{n},t_{0}),\nonumber \\
\\
{\cal D}(t,t_{0})&=&e^{i\int^{t}_{t_{0}}ds LQ(s)}_{\rightarrow}
(U^{Q}_{0})^{-1}(t,t_{0}) \nonumber \\
&=& 1+\sum^{\infty}_{n=1}i^n 
\int^{t}_{t_{0}}dt_{1}\int^{t_{1}}_{t_{0}}dt_{2}
\cdots\int^{t_{n-1}}_{t_{0}}dt_{n}
\breve{L}^{Q}_{I}(t_{1},t_{0})\breve{L}^{Q}_{I}(t_{2},t_{0})\cdots
\breve{L}^{Q}_{I}(t_{n},t_{0}), \nonumber \\
\end{eqnarray}
where
\begin{eqnarray}
\breve{L}_{I}(t,t_{0}) &=& U_{0}(t,t_{0})L_{I}U^{-1}_{0}(t,t_{0}), \\
\breve{L}^{Q}_{I}(t,t_{0}) &=& U^{Q}_{0}(t,t_{0})L_{I}Q(U^{Q}_{0})^{-1}(t,t_{0}), \\
U_{0}(t,t_{0}) &=& e^{i\int^{t}_{t_{0}}ds L_{0}(s,t_{0})}_{\rightarrow}, \\
U^{Q}_{0}(t,t_{0}) &=& e^{i\int^{t}_{t_{0}}ds
L_{0}(s,t_{0})Q(s)}_{\rightarrow}.
\end{eqnarray}
It is noteworthy that 
the term including the time derivative of the projection operator 
disappears.
The operator ${\cal C}(t,t_{0})$ [${\cal D}(t,t_{0})$] 
is a time [an anti-time] 
ordered function of Liouville operators.
(The details of these expressions are given in Appendix \ref{app:3}.)
Then, we have 
\begin{eqnarray}
P(t)\Sigma(t,t_{0})\frac{1}{1-\Sigma(t,t_{0})}
&=& -P(t)U^{-1}_{0}(t,t_{0}){\cal C}(t,t_{0})Q(t)
\frac{1}{1+({\cal C}(t,t_{0})-1)Q(t)}U_{0}(t,t_{0}).\nonumber \\
\end{eqnarray}
To derive this expression, we have used mathematical induction.
\cite{ref:Koi-Mar}
(The detailed derivation appears in Appendix \ref{app:4}.)
Substituting the above result into Eq.~(\ref{eqn:TCL}), we obtain 
\begin{eqnarray}
\frac{d}{dt}O(t) &=& e^{iL(t-t_{0})}P(t)iLO(t_{0}) \nonumber \\
&&+e^{iL(t-t_{0})}P(t)U^{-1}_{0}(t,t_{0}){\cal C}(t,t_{0})Q(t)
\frac{1}{1-({\cal C}(t,t_{0})-1)Q(t)}U_{0}(t,t_{0})iLO(t_{0}) \nonumber \\
&&+Q(t_{0})e^{i\int^{t}_{t_{0}}ds LQ(s)}_{\rightarrow}
\frac{1}{1-\Sigma(t,t_{0})}iLO(t_{0}).
\end{eqnarray}
This form of the TCL equation is convenient to 
carry out the perturbative expansion.

When we expand $P\Sigma(t,t_{0})/(1-\Sigma(t,t_{0}))$ 
up to first order in the 
interaction $H_{I}$, we have
\begin{eqnarray}
\lefteqn{\frac{d}{dt}O(t)} && \nonumber \\
&=& e^{iL(t-t_{0})}P(t)iLO(t_{0})\nonumber \\
&& -e^{iL(t-t_{0})}P(t)U^{-1}_{0}(t,t_{0})Q(t)U_{0}(t,t_{0})iLO(t_{0})
\nonumber\\
&& +e^{iL(t-t_{0})}P(t)U^{-1}_{0}(t,t_{0})\int^{t}_{t_{0}}ds
U_{0}(s,t_{0})iL_{I}(s,t_{0})U^{-1}_{0}(s,t_{0})Q(t)U_{0}(t,t_{0})iLO(t_{0})
\nonumber\\
&& -e^{iL(t-t_{0})}P(t)U^{-1}_{0}(t,t_{0})Q(t)\nonumber \\
&&\times \int^{t}_{t_{0}}ds
U_{0}(s,t_{0})iL_{I}(s,t_{0})U^{-1}_{0}(s,t_{0})
Q(t)U_{0}(t,t_{0})iLO(t_{0})
\nonumber \\
&&+Q(t_{0})e^{i\int^{t}_{t_{0}}ds LQ(s)}_{\rightarrow}
\frac{1}{1-\Sigma(t,t_{0})}iLO(t_{0}) \nonumber \\
&=& e^{iL(t-t_{0})}P(t)U^{-1}_{0}(t,t_{0})P(t)U_{0}(t,t_{0})iLO(t_{0}) 
\nonumber \\
&& +e^{iL(t-t_{0})}P(t)U^{-1}_{0}(t,t_{0})P(t)\int^{t}_{t_{0}}
ds U_{0}(s,t_{0})iL_{I}U^{-1}_{0}(s,t_{0})Q(t)U_{0}(t,t_{0})iLO(t_{0}) \nonumber \\
&& +Q(t_{0})e^{i\int^{t}_{t_{0}}ds LQ(s)}_{\rightarrow}
\frac{1}{1-\Sigma(t,t_{0})}iLO(t_{0}). \label{eqn:TTCL}
\end{eqnarray}
It is easily seen that Eq.~(\ref{eqn:TTCL}) does not contain 
a time-convolution integral.
If this did contain a time-convolution integral, 
the form of the full time-evolution operator, $e^{iL(t-t_{0})}$, 
which operates from the left in the second term on the r.h.s. 
of this equation, must be $e^{iL(t-s)}$, where $s$ is an integral variable.
\cite{ref:HSS,ref:Koi-Mar}
Therefore, this equation is called the TCL equation.

When we ignore the time dependence of the projection operator, 
Eq.~(\ref{eqn:TTCL}) is the same as Eq.~(2$\cdot$31) 
in Ref.~\cite{ref:Koi-Mar}.
It should be noted that not $Q(t)$ but $Q(t_{0})$ 
operates from the left in the third term on the r.h.s. of the equation.
Hence, this term represents the effect of the initial correlation 
of the initial density matrix.
This point becomes more clear in the next section.
In the usual projection operator method, 
such a term is interpreted as the fluctuation force.
Therefore, the third term can be regarded as the fluctuation force 
also in this time-dependent projection operator method.

\section{Coupled transport equations}\label{sec:3}

Now, we apply the time-dependent projection operator method 
to a quantum field theoretical model and derive a transport equation.
We consider the Hamiltonian with two boson fields  
\begin{eqnarray}
H &=& H_{\sigma} + H_{\pi} + H_{I}, 
\end{eqnarray}
where
\begin{eqnarray}
H_{\sigma} &=& \int d^3 {\bf x} 
\frac{1}{2}\{ \Phi^2 (x) + (\nabla \phi(x))^2 + m^2_{\sigma} \phi^2
(x)\}, \\
H_{\pi}  &=& \int d^3 {\bf x} \frac{1}{2}\{ \Pi^2 (x) + (\nabla
\pi(x))^2 + m^2_{\pi} \pi^2 (x)\}, \\
H_{I}   &=& \int d^3 {\bf x} g \pi^2 (x)\phi(x).
\end{eqnarray}
Here, $\Phi(x)$ and $\Pi(x)$ are the conjugate fields of $\phi(x)$ and 
$\pi(x)$, respectively.
We call the particle represented by the $\phi(x)$ field 
the $\sigma$ boson, and that represented by the $\pi(x)$ field the $\pi$ boson.
The nonperturbative and interaction Hamiltonians are given by
$H_{0}=H_{\sigma}+H_{\pi}$ and $H_{I}$, respectively.
Here, we ignore the time dependence of the nonperturbative Hamiltonian 
for simplicity.

These four fields are expanded as 
\begin{eqnarray}
\phi({\bf x},t_{0}) 
&=& \sum_{\bf k}\frac{1}{\sqrt{2V\omega^{\sigma}_{\bf k}}}
(a_{\bf k}(t_{0})e^{i{\bf kx}} + a^{\dagger}_{\bf k}(t_{0})e^{-i{\bf kx}}), \\
\Phi({\bf x},t_{0})
&=& -i\sum_{\bf k}\sqrt{\frac{\omega^{\sigma}_{\bf k}}{2V}}
(a_{\bf k}(t_{0})e^{i{\bf kx}} - a^{\dagger}_{\bf k}(t_{0})e^{-i{\bf kx}}), \\
\pi({\bf x},t_{0}) 
&=& \sum_{\bf k}\frac{1}{\sqrt{2V\omega^{\pi}_{\bf k}}}
(b_{\bf k}(t_{0})e^{i{\bf kx}} + b^{\dagger}_{\bf k}(t_{0})e^{-i{\bf kx}}), \\
\Pi({\bf x},t_{0}) 
&=& -i\sum_{\bf k}\sqrt{\frac{\omega^{\pi}_{\bf k}}{2V}}
(b_{\bf k}(t_{0})e^{i{\bf kx}} - b^{\dagger}_{\bf k}(t_{0})e^{-i{\bf kx}}),
\end{eqnarray}
where 
$\omega^{\sigma}_{\bf k} = \sqrt{{\bf k}^2 + m_{\sigma}^2}$ and 
$\omega^{\pi}_{\bf k} = \sqrt{{\bf k}^2 + m_{\pi}^2}$.
Here, $V$ and $t_{0}$ are the volume of the total system and 
the initial time at which we prepare 
the initial state.
We take the limit $V \longrightarrow \infty$ at the end of the 
calculation.
The creation and annihilation operators 
of the $\sigma$ and $\pi$ particles 
are subject to the following commutation relations:
\begin{eqnarray}
[a_{\bf k}(t_{0}),a^{\dagger}_{\bf k'}(t_{0})]
=
[b_{\bf k}(t_{0}),b^{\dagger}_{\bf k'}(t_{0})] 
= \delta^{(3)}_{\bf k,k'}.
\end{eqnarray}
Here, $[~~]$ represents the commutator.
All other commutators vanish.

For simplicity, 
we consider the case in which the initial density matrix is given by the 
direct product of the $\sigma$ boson and the $\pi$ boson density
matrices:
\begin{eqnarray}
\rho_{0} = \rho_{0 \sigma}\otimes\rho_{0 \pi}.  \label{eqn:ini-den}
\end{eqnarray}
This means that there is no initial correlation between the $\sigma$ boson 
and the $\pi$ boson.

The system we are considering consists of two degrees of freedom, 
that of the $\sigma$ boson and that of the $\pi$ boson.
In the usual time-independent projection operator method, 
it is often assumed that one degree of freedom plays the role of 
a heat bath in thermal equilibrium 
for another degree of freedom.\cite{ref:Reno}
In the present application, we treat the two degrees of freedom
on an equal footing and 
derive coupled transport equations.
\cite{ref:Will-Pic,ref:Gra-Wei,ref:Shi-Has-tp}
First, we calculate the transport equation of the $\sigma$ boson.
In this case, the $\sigma$ boson is regarded as the system and 
the $\pi$ boson as the environment that is to be coarse-grained, 
respectively.
To integrate out the environment degree of freedom, 
we define the time-dependent projection operator as
\begin{eqnarray}
P(t) O &=& {\rm Tr}_{E}[\rho_{E}(t) O], \label{eqn:TDP}
\end{eqnarray}
where
\begin{eqnarray}
\rho_{E}(t) &=& e^{-\beta(t)H_{\pi}}/Z_{\pi}(t), \label{eqn:led} \\
Z_{\pi}(t) &=& {\rm Tr}[e^{-\beta(t)H_{\pi}}].
\end{eqnarray}
Here, $\rho_{E}(t)$ is the local equilibrium density matrix.
This projection operator 
satisfies the conditions (\ref{eqn:PPP}) and (\ref{eqn:QL0P}).
This choice of $\rho_{E}(t)$ 
comes from our implicit assumption that 
the time evolution of the $\pi$ boson 
is well approximated by the local equilibrium density matrix with 
the time-dependent temperature $\beta^{-1}(t)$.
It is possible to calculate the transport equation 
without assuming local equilibrium in
the time evolution of the system.
However, in such a case, 
it is necessary to calculate a large number of correlation functions.
To avoid this difficulty, we assume local equilibrium in this paper.
Now, we employ the following initial condition: 
\begin{eqnarray}
\rho_{E}(t_{0}) = \rho_{0\pi}. \label{eqn:IPC}
\end{eqnarray}
This condition is needed to eliminate the contribution from the third term 
on the r.h.s. of Eq.~(\ref{eqn:TTCL}).
This becomes clear in the next paragraph.
The time dependence of the temperature is determined later.

Substituting $O(t_{0})=a^{\dagger}_{\bf k}(t_{0})a_{\bf k}(t_{0})$ 
into Eq.~(\ref{eqn:TCL}) and using the definition (\ref{eqn:TDP}), we obtain 
\begin{eqnarray}
\lefteqn{\frac{d}{dt}a^{\dagger}_{\bf k}(t)a_{\bf k}(t)} && \nonumber \\
&=& \frac{ig}{\sqrt{2V\omega^{\sigma}_{\bf k}}}
\langle \rho_{\bf k}(t,t_{0}) \rangle_{t} (a_{\bf k}(t) - a^{\dagger}_{\bf k}(t)) \nonumber \\
&&+\frac{g^2}{4V\omega^{\sigma}_{\bf k}}\int^{t}_{t_{0}}ds 
\{
\langle [\rho_{\bf -k}(s,t_{0}),\rho_{\bf k}(t,t_{0})]_{+} \rangle_{t} 
e^{i\omega^{\sigma}_{\bf k}(s-t)} 
+ \langle [\rho_{\bf k}(s,t_{0}),\rho_{\bf -k}(t,t_{0})]_{+} \rangle_{t}
e^{-i\omega^{\sigma}_{\bf k}(s-t)} \nonumber \\
&&-2\langle \rho_{\bf -k}(s,t_{0}) \rangle_{t} \langle \rho_{\bf k}(t,t_{0}) \rangle_{t}
e^{i\omega^{\sigma}_{\bf k}(s-t)} 
-2\langle \rho_{\bf -k}(s,t_{0}) \rangle_{t} \langle \rho_{\bf k}(t,t_{0}) \rangle_{t}
e^{-i\omega^{\sigma}_{\bf k}(s-t)} 
\} \nonumber \\
&& -\frac{g^2}{4V\omega^{\sigma}_{\bf k}}\int^{t}_{t_{0}}ds
\{
\langle [\rho_{\bf -k}(s,t_{0}),\rho_{\bf k}(t,t_{0})] \rangle_{t}
[a_{\bf k}(t), a_{\bf -k}(t)e^{-i\omega^{\sigma}_{\bf k}(s-t)} 
+a_{\bf k}^{\dagger}(t)e^{i\omega^{\sigma}_{\bf k}(s-t)} ]_{+} \nonumber \\
&& - \langle [\rho_{\bf k}(s,t_{0}),\rho_{\bf -k}(t,t_{0})] \rangle_{t} 
[a^{\dagger}_{\bf k}(t), a_{\bf k}(t)e^{-i\omega^{\sigma}_{\bf k}(s-t)} 
+a_{\bf -k}^{\dagger}(t)e^{i\omega^{\sigma}_{\bf k}(s-t)} ]_{+}
\} \nonumber \\
&& + flu,  \label{eqn:s-eq}
\end{eqnarray}
where
\begin{eqnarray}
\langle \rho_{\bf k}(t,t_{0}) \rangle_{t} 
&=& \int d^3 {\bf x} e^{i{\bf kx}}{\rm Tr}[\rho_{E}(t)\pi^2 ({\bf x},t;t_{0})], \label{eqn:rhop}\\
\langle [\rho_{\bf k_{1}}(s,t_{0}),\rho_{\bf k_{2}}(t,t_{0})] \rangle_{t}
&=& \int d^3 {\bf x}_{1}d^3 {\bf x}_{2}e^{i{\bf k_{1}x_{1}}}e^{i{\bf k_{2}x_{2}}}
{\rm Tr}[ \rho_{E}(t)[\pi^2 ({\bf x}_{1},s;t_{0}),\pi^2 ({\bf
x}_{2},t;t_{0})] ]. \nonumber \\
\label{eqn:rhop2}\\
\langle [\rho_{\bf k_{1}}(s,t_{0}),\rho_{\bf k_{2}}(t,t_{0})]_{+} \rangle_{t}
&=& \int d^3 {\bf x}_{1}d^3 {\bf x}_{2}e^{i{\bf k_{1}x_{1}}}e^{i{\bf k_{2}x_{2}}}
{\rm Tr}[ \rho_{E}(t)[\pi^2 ({\bf x}_{1},s;t_{0}),\pi^2 ({\bf
x}_{2},t;t_{0})]_{+} ]. \nonumber \\
\label{eqn:rhop3}
\end{eqnarray}
Here, $[~~]_{+}$ is the anti-commutator.
The last term, $flu$, on the r.h.s. of Eq.~(\ref{eqn:s-eq}) 
comes from the third term on the r.h.s. of Eq.~(\ref{eqn:TTCL}).
When we calculate the expectation value by using the initial density matrix, 
this term vanishes because of the definition of $flu$ 
and the condition (\ref{eqn:IPC}).
When we consider the initial density matrix with an initial correlation, 
we cannot employ the condition (\ref{eqn:ini-den}), 
and the third term has a finite value.
In short, the third term represents the effect of the initial
correlation.
We can thus regard the third term $flu$ as the fluctuation force term
as in the usual projection operator method.
In this calculation, we do not consider the initial 
correlation, and therefore, we ignore this term.

Before continuing with the calculation, we discuss the structure of 
this equation.
This equation has strange terms that are proportional to 
$a^{\dagger}_{\bf k}(t)$, $a_{\bf k}(t)$, 
$a^{\dagger}_{\bf k}(t)a^{\dagger}_{\bf-k}(t)$ 
and $a_{\bf k}(t)a_{\bf-k}(t)$.
Such terms are not seen in the usual Boltzmann equation, which is 
constituted of terms proportional 
to $a^{\dagger}_{\bf k}(t)a_{\bf k}(t)$.
The terms proportional to $a^{\dagger}_{\bf k}(t)$ and $a_{\bf k}(t)$ 
survive when we consider the case that there is a condensate of 
the $\sigma$ boson.
Here, we do not consider the case of the condensate, 
and therefore we ignore such terms. 
The existence of the terms 
proportional to $a^{\dagger}_{\bf k}(t)a^{\dagger}_{\bf-k}(t)$ 
and $a_{\bf k}(t)a_{\bf-k}(t)$ was 
pointed out in Ref.~\cite{ref:SHH}.
In that paper it is noted that the existence of such terms is homologous to 
a parametric amplifier, and 
$a^{\dagger}_{\bf k}(t)a^{\dagger}_{\bf-k}(t)$, 
$a_{\bf k}(t)a_{\bf-k}(t)$ and $a^{\dagger}_{\bf k}(t)a_{\bf k}(t)$ 
form the {\it SU}(1,1) symmetry group.
In this calculation, we drop such terms for simplicity.

To calculate the correlation functions (\ref{eqn:rhop}), 
(\ref{eqn:rhop2}) and (\ref{eqn:rhop3}), 
it is convenient to use a technique of thermo-field dynamics (TFD).
(The detailed calculation is given in Appendix \ref{app:5}.)
Simply quoting the result, we obtain 
\begin{eqnarray}
{\rm Tr}[\rho_{E}(t) \rho_{\bf k}(t,t_{0})]
&=& \int d^3 {\bf x}e^{i{\bf kx}}
\ccontra{\pi({\bf x},t)}{\pi({\bf x},t)}, \\
{\rm Tr}[\rho_{E}(t) [\rho_{\bf k_{1}}(s,t_{0}),\rho_{\bf k_{2}}(t,t_{0})]]
&=& \int d^3 {\bf x}_{1}d^3 {\bf x}_{2}
e^{i{\bf k}_{1}{\bf x}_{1}}e^{i{\bf k}_{2}{\bf x}_{2}} \nonumber \\
&& \times 2 \{
(\ccontrad{\pi({\bf x}_{1},s)}{\pi({\bf x}_{2},t)}
\hspace{1.2cm}
)^2
-(\ccontrad{\pi({\bf x}_{2},t)}{\pi({\bf x}_{1},s)}
\hspace{1.3cm}
)^2
\},\nonumber \\
\end{eqnarray}
where
\begin{eqnarray}
\ccontradd{\pi({\bf x}_{1},t_{1})}{\pi({\bf x}_{2},t_{2})}
\hspace{1.3cm}
&=& \sum_{\bf k}\frac{1}{2V\omega^{\pi}_{\bf k}}
(
(1+n^{\pi}_{\bf k}(t))e^{-i\omega^{\pi}_{\bf k}(t_{1}-t_{2})}
+n^{\pi}_{\bf k}(t) e^{i\omega^{\pi}_{\bf k}(t_{1}-t_{2})}
)e^{i{\bf k}({\bf x}_{1}-{\bf x}_{2})}.\nonumber \\
\end{eqnarray}

Substituting the above results into Eq.~(\ref{eqn:s-eq}) 
and taking the expectation value with
respect to the initial density matrix, 
we obtain the transport equation of the $\sigma$ distribution function, 
\begin{eqnarray}
\frac{d}{dt}n^{\sigma}_{\bf k}
&=& \frac{g^2}{2(2\pi)^3 \omega^{\sigma}_{\bf k}} 
\int^{t}_{t_{0}} ds \int d^3 
{\bf q} 
\frac{1}{\omega^{\pi}_{\bf q} \omega^{\pi}_{\bf q+k}} \nonumber \\
&& \times [\{(1+n^{\pi}_{\bf q})(1+n^{\pi}_{\bf q+k})(1+n^{\sigma}_{\bf k})
-n^{\pi}_{\bf q}n^{\pi}_{\bf q+k}n^{\sigma}_{\bf k}\}
\cos (\omega^{\pi}_{\bf q}+\omega^{\pi}_{\bf q+k}
+\omega^{\sigma}_{\bf k})(s-t) \nonumber \\
&&+\{n^{\pi}_{\bf q}n^{\pi}_{\bf q+k}(1+n^{\sigma}_{\bf k})
-(1+n^{\pi}_{\bf q})(1+n^{\pi}_{\bf q+k})n^{\sigma}_{\bf k}
\}
\cos (\omega^{\pi}_{\bf q}+\omega^{\pi}_{\bf q+k}
-\omega^{\sigma}_{\bf k})(s-t) \nonumber \\
&&+\{(1+n^{\pi}_{\bf q})n^{\pi}_{\bf q+k}(1+n^{\sigma}_{\bf k})
-n^{\pi}_{\bf q}(1+n^{\pi}_{\bf q+k})n^{\sigma}_{\bf k}\} \cos (\omega^{\pi}_{\bf q}-\omega^{\pi}_{\bf q+k}+\omega^{\sigma}_{\bf
k})(s-t) \nonumber \\
&&+\{n^{\pi}_{\bf q}(1+n^{\pi}_{\bf q+k})(1+n^{\sigma}_{\bf k})
-(1+n^{\pi}_{\bf q})n^{\pi}_{\bf q+k}n^{\sigma}_{\bf k}\}
\cos (\omega^{\pi}_{\bf q}-\omega^{\pi}_{\bf q+k}-\omega^{\sigma}_{\bf
k})(s-t) ], \nonumber \\
\label{eqn:Sigma-n2}
\end{eqnarray}
where $n^{\sigma}_{\bf k}(t) 
= {\rm Tr}[\rho a^{\dagger}_{\bf k}(t_{0})a_{\bf k}(t_{0})]$.
Here we have omitted the time dependence of $n^{\sigma}_{\bf k}$ 
and $n^{\pi}_{\bf k}$ for simplicity.
Each term in the braces on the r.h.s. of this equation 
is composed of two contributions: the gain term and the loss term. 
The first term describes the creation of two $\pi$ bosons and one 
$\sigma$ boson minus their annihilation.
The second term describes the creation of one $\sigma$ boson and 
the annihilation of two $\pi$ bosons 
minus the annihilation of one $\sigma$ boson 
and the creation of two $\pi$ bosons.
The third and fourth terms describe the creation of 
one $\sigma$ boson and one $\pi$ boson and the annihilation of 
one $\pi$ boson minus the annihilation 
of one $\sigma$ boson and one $\pi$ boson and the creation 
of one $\pi$ boson.
This is a reasonable result, which is expected from the usual 
Boltzmann equation.\cite{ref:NETFD,ref:Kad-Bay}
When we take the limit $t_{0}\longrightarrow -\infty$, 
Dirac delta functions 
that preserve energy conservation are obtained.

To solve the above transport equation, it is necessary to 
determine the time dependence of the $\pi$ distribution function.
Our next task is to obtain the transport equation for the $\pi$ boson.
Our strategy is to treat the $\sigma$ and $\pi$ degrees of freedom 
on an equal footing.
We thus reverse the roles of the $\sigma$ and $\pi$ degrees of freedom;
that is, we regard the $\pi$ boson as the system and the $\sigma$ boson as 
the environment and 
calculate the transport equation of the $\pi$ boson using the same procedure 
as in the calculation of Eq.~(\ref{eqn:Sigma-n2}).
In this way, we obtain the coupled transport equations of 
the $\sigma$ and $\pi$ distribution functions.

The projection operator is given by
\begin{eqnarray}
P(t)O &=& {\rm Tr}_{E}[\rho_{E}(t)O], 
\end{eqnarray}
where
\begin{eqnarray}
\rho_{E}(t) &=& e^{-\beta(t)H_{\sigma}}/Z_{\phi}(t), \label{eqn:re-sig}\\
Z_{\phi}(t) &=& {\rm Tr}[e^{-\beta(t)H_{\sigma}}].
\end{eqnarray}
As in the case of the $\sigma$ boson, the $\sigma$ distribution function 
is included in the $\pi$ transport equation.
We assume local equilibrium in the time evolution again, i.e., 
\begin{eqnarray}
{\rm Tr}[\rho_{E}(t) a^{\dagger}_{\bf k}(t_{0})a_{\bf k}(t_{0}) ]
= n^{\sigma}_{\bf k}(t),
\end{eqnarray}
where $n^{\sigma}_{\bf k}(t)$ is given by the solution of 
Eq.~(\ref{eqn:Sigma-n2})

When we use the definition (\ref{eqn:re-sig}) 
and substitute $O(t_{0})=b^{\dagger}(t_{0})b(t_{0})$ into Eq.~(\ref{eqn:TCL}), 
we can obtain the transport equation of the $\pi$ distribution function.
In this case, the transport equation has fourth order 
correlation functions of the $\pi$ boson. 
To obtain a closed form for the coupled transport equations, 
we approximate such terms as
\begin{eqnarray}
{\rm Tr}[\rho_{0} b^{\dagger}_{\bf k}(t)b^{\dagger}_{\bf l}(t)b_{\bf m}(t)b_{\bf n}(t)]
= n^{\pi}_{\bf k}(t)n^{\pi}_{\bf l}(t)
\delta^{(3)}_{\bf k,m}\delta^{(3)}_{\bf l,n} 
+ n^{\pi}_{\bf k}(t)n^{\pi}_{\bf l}(t)
\delta^{(3)}_{\bf k,n}\delta^{(3)}_{\bf l,m}. \nonumber \\
\end{eqnarray}

Finally, the transport equation for the $\pi$ distribution function is
\begin{eqnarray}
\frac{d}{dt}n^{\pi}_{\bf k} 
&=& 
\int^{t}_{t_{0}}ds \int d^3 {\bf l}
\frac{g^2}{(2\pi)^3 \omega^{\pi}_{\bf k}\omega^{\pi}_{\bf
l}\omega^{\sigma}_{\bf l+k}} \nonumber \\
&&\times [ \{ (1+n^{\sigma}_{\bf k+l})
(1+n^{\pi}_{\bf k})(1+n^{\pi}_{\bf l})
-n^{\sigma}_{\bf k+l}n^{\pi}_{\bf k}n^{\pi}_{\bf l}
\}\cos(\omega^{\sigma}_{\bf l+k}+\omega^{\pi}_{\bf l} +
\omega^{\pi}_{\bf k})(s-t) \nonumber \\
&& + 
\{ 
n^{\sigma}_{\bf k+l}(1+n^{\pi}_{\bf k})(1+n^{\pi}_{\bf l})
-(1+ n^{\sigma}_{\bf k+l})n^{\pi}_{\bf k}n^{\pi}_{\bf l}
\}\cos(\omega^{\sigma}_{\bf l+k}-\omega^{\pi}_{\bf l} -
\omega^{\pi}_{\bf k} )(s-t) \nonumber\\
&&+  
\{
(1+n^{\sigma}_{\bf k+l})(1+n^{\pi}_{\bf k})n^{\pi}_{\bf l}
-n^{\sigma}_{\bf k+l}(1+n^{\pi}_{\bf l})n^{\pi}_{\bf k}
\}\cos(\omega^{\sigma}_{\bf l+k}-\omega^{\pi}_{\bf l} +
\omega^{\pi}_{\bf k} )(s-t) \nonumber \\
&&+  
\{
n^{\sigma}_{\bf k+l}(1+n^{\pi}_{\bf k})n^{\pi}_{\bf l}
-(1+n^{\sigma}_{\bf k+l})(1+n^{\pi}_{\bf l})n^{\pi}_{\bf k}
\}\cos(\omega^{\sigma}_{\bf l+k}+\omega^{\pi}_{\bf l} -
\omega^{\pi}_{\bf k} )(s-t)]. \nonumber \\
\label{eqn:p-eq}
\end{eqnarray}

Each term in the braces on the r.h.s. of Eq.~(\ref{eqn:p-eq}) 
can be interpreted as gain minus loss processes, as 
in the case of the $\sigma$ transport equation.
The time-evolution of the $\sigma$ and $\pi$ distribution functions 
are obtained by solving this coupled set of equations.

\section{Summary and conclusions}\label{sec:4}

We have derived a systematic perturbative expansion formula 
for a Langevin-type equation without time-convolution integral terms.
In this formalism, the irrelevant subsystem can have 
some time dependence, 
and therefore, we can treat 
a more general nonequilibrium process in which the mass, the temperature, 
and so on, are time dependent.
When we ignore the time dependence of the projection operator, 
we can reproduce the result of the usual projection 
operator method.\cite{ref:Koi-Mar}
Furthermore, the third term on the r.h.s. of Eq.(\ref{eqn:TTCL}) 
represents the effect of the initial correlation.
We can thus interpret it as the fluctuation force even in the 
time-dependent projection operator method.

We applied this formalism to a quantum field theoretical 
model that consists of $\sigma$ and $\pi$ bosons and thereby 
obtained coupled transport equations for the two bosons.
Deriving these coupled transport equations, 
the correlation functions were calculated with respect to 
the local equilibrium density matrix.
In other words, we have assumed 
that the time evolution of the system 
can be well approximated by the local equilibrium distribution function.
The derived equations have 
terms that are not seen in the usual Boltzmann equation.
It is worth studying the effect of such terms on the 
transport equation, but we have dropped them in this paper.
Each transport equation has a form that can be interpreted as
gain minus loss processes, which is the structure usually seen in 
the Boltzmann equation.
This gives reason to believe that the time-dependent projection operator 
method developed in this paper is a valid formalism 
to describe nonequilibrium processes.

In this paper, we have ignored the time dependence 
of the nonperturbative Hamiltonian.
In general nonequilibrium processes, it is possible 
to consider the situation 
in which the mass of a particle changes with time.
In this case, the time dependence of the nonperturbative Hamiltonian 
is caused by the time-dependent mass.
We will report on an investigation of this effect in a future publication.

Finally, we mention the fluctuation-dissipation theorem.
This is a famous theorem in statistical mechanics that expresses 
the relation between macroscopic transport coefficients 
and microscopic fluctuations.
It is well-known that there are two expressions for 
the fluctuation-dissipation theorem.
For the Langevin equation, 
the fluctuation-dissipation theorem of the second kind is important.
This theorem was first proven by Mori using a time-independent 
projection operator technique.\cite{ref:Mori}
Furukawa developed the time-dependent Mori projection operator, according to 
which the generalized version of 
the fluctuation-dissipation theorem of the second kind 
was obtained.\cite{ref:Furu}
The time-dependent Mori projection operator, which satisfies $P(t)P(t')=P(t)$, 
is not included in our formalism, due to condition (\ref{eqn:PPP}).
Our formalism includes the time-independent projection operator 
method and reproduces the Tokuyama-Mori equation\cite{ref:Toku-Mori} 
when we use the Mori projection operator.
Therefore, the fluctuation force in our formalism agrees with 
that in the Tokuyama-Mori formalism.
We thus conclude that our fluctuation force is 
a natural extension of theirs, which satisfies 
the fluctuation-dissipation theorem.
However, the fluctuation-dissipation theorem 
in our formalism has still not been established, 
and demonstrating it is one important future problem.

\begin{center}
{\bf ACKNOWLEDGMENTS}
\end{center}

The author thanks F.~Shibata and Y.~Yamanaka for useful discussions.

\appendix

\section{Definition of the Operators ${\cal C}$ and ${\cal D}$}\label{app:3}

The definitions of the operators ${\cal C}(t,t_{0})$ and ${\cal D}(t,t_{0})$ 
are
\begin{eqnarray}
{\cal C}(t,t_{0}) &=& U_{0}(t,t_{0})
e^{-iL(t-t_{0})}, \\
{\cal D}(t,t_{0}) &=& e^{i\int^{t}_{t_{0}}ds LQ(s)}_{\rightarrow}
(U^{Q}_{0})^{-1}(t,t_{0}),
\end{eqnarray}
where
\begin{eqnarray}
U_{0}(t,t_{0}) 
&=& e^{i\int^{t}_{t_{0}}dsL_{0}(s,t_{0})}_{\rightarrow}, \\
U^{Q}_{0}(t,t_{0}) 
&=& e^{i\int^{t}_{t_{0}}ds L_{0}(s,t_{0})Q(s)}_{\rightarrow}.
\end{eqnarray}
These operators must satisfy the following differential equations:
\begin{eqnarray}
\frac{d}{dt}{\cal C}(t,t_{0}) 
&=& U_{0}(t,t_{0})(iL_{0}(t,t_{0})-iL)e^{-iL(t-t_{0})} \nonumber \\
&=& -i\breve{L}_{I}(t,t_{0}){\cal C}(t,t_{0}), \\
\frac{d}{dt}{\cal D}(t,t_{0}) 
&=& e^{i\int^{t}_{t_{0}}ds LQ(s)}_{\rightarrow}(iL-iL_{0}(t,t_{0}))Q(t)
(U^{Q}_{0})^{-1}(t,t_{0}) \nonumber \\
&=& {\cal D}(t,t_{0})i\breve{L}^{Q}_{I}(t,t_{0}),
\end{eqnarray}
where 
\begin{eqnarray}
\breve{L}_{I}(t,t_{0}) &=& U_{0}(t,t_{0})L_{I}(t,t_{0})U^{-1}_{0}(t,t_{0}), \\
\breve{L}^{Q}_{I}(t,t_{0}) &=& U^{Q}_{0}(t,t_{0})L_{I}(t,t_{0})
Q(U^{Q}_{0})^{-1}(t,t_{0}).
\end{eqnarray}
The solutions of the above differential equations are 
\begin{eqnarray}
{\cal C}(t,t_{0}) &=& 
1+\sum^{\infty}_{n=1}(-i)^n 
\int^{t}_{t_{0}}dt_{1}\int^{t_{1}}_{t_{0}}dt_{2}
\cdots\int^{t_{n-1}}_{t_{0}}dt_{n}
\breve{L}_{I}(t_{1},t_{0})\breve{L}_{I}(t_{2},t_{0})\cdots\breve{L}_{I}(t_{n},t_{0}),
\nonumber \\
\\
{\cal D}(t,t_{0})
&=& 1+\sum^{\infty}_{n=1}i^n 
\int^{t}_{t_{0}}dt_{1}\int^{t_{1}}_{t_{0}}dt_{2}
\cdots\int^{t_{n-1}}_{t_{0}}dt_{n}
\breve{L}^{Q}_{I}(t_{1},t_{0})\breve{L}^{Q}_{I}(t_{2},t_{0})\cdots
\breve{L}^{Q}_{I}(t_{n},t_{0}).\nonumber \\
\end{eqnarray}

\section{Transformation of the Operator $\Sigma(t,t_{0})$}\label{app:4}

The operator $\Sigma(t,t_{0})$ can be rewritten as 
\begin{eqnarray}
\Sigma(t,t_{0}) 
&=& \int^{t}_{t_{0}}ds e^{-iL(t-s)} 
\{
\dot{Q}(s)+P(s)iLQ(s)
\} e^{i\int^{t}_{s}d\tau LQ(\tau)}_{\rightarrow} \nonumber \\
&=& \int^{t}_{t_{0}}ds e^{-iL(t-s)}
\frac{d}{ds}( Q(s)e^{i\int^{t}_{s}d\tau LQ(\tau)}_{\rightarrow} )
-\int^{t}_{t_{0}}ds e^{-iL(t-s)}\frac{d}{ds}e^{i\int^{t}_{s}d\tau LQ(\tau)}_{\rightarrow}  \nonumber \\
&=& -P(t) + e^{-iL(t-t_{0})}P(t_{0})e^{i\int^{t}_{t_{0}}d\tau LQ(\tau)}_{\rightarrow}
+\int^{t}_{t_{0}}ds e^{-iL(t-s)}iLP(s)e^{i\int^{t}_{s}d\tau LQ(\tau)}_{\rightarrow}   \nonumber \\
&=& -P(t) + e^{-iL(t-t_{0})}P(t_{0})e^{i\int^{t}_{t_{0}}d\tau LQ(\tau)}_{\rightarrow}
+\int^{t}_{t_{0}}ds \frac{d}{ds} e^{-iL(t-s)}e^{i\int^{t}_{s}d\tau LQ(\tau)}_{\rightarrow}   \nonumber \\
&=& Q(t) - e^{-iL(t-t_{0})}
Q(t_{0})e^{i\int^{t}_{t_{0}}d\tau LQ(\tau)}_{\rightarrow} \nonumber \\
&=& Q(t) - U_{0}^{-1}(t,t_{0}){\cal C}(t,t_{0})Q(t_{0})
{\cal D}(t,t_{0})e^{i\int^{t}_{t_{0}}ds L_{0}(s,t_{0})Q(s)}_{\rightarrow}.
\end{eqnarray}
Using mathematical induction, we confirm the following relation:
\begin{eqnarray}
\lefteqn{P(t)\Sigma(t,t_{0})(Q(t)\Sigma(t,t_{0}))^n}&& \nonumber \\
&=& [ (-1)^{n-1}P(t)\{ \tilde{Q}(t)(\tilde{{\cal C}}(t)-1) \}^n \tilde{Q}(t)
     +(-1)^{n-1}P(t)\{ (\tilde{{\cal C}}(t) -1) \tilde{Q}(t)\}^{n+1}] \nonumber \\
&& + P(t)\sum_{l=0}^{n-1}(-1)^l \{ \tilde{Q}(t)(\tilde{{\cal C}}(t)-1) \}^l \tilde{Q}(t)(\tilde{{\cal D}}(t)-1)
     (\tilde{Q}(t)\Sigma(t,t_{0}))^{n-1-l} \nonumber \\
&& + P(t)\sum_{l=0}^{n-1}(-1)^l \{ (\tilde{{\cal C}}(t)-1)\tilde{Q}(t) \}^{l+1} (\tilde{{\cal D}}(t)-1)
     (\tilde{Q}(t)\Sigma(t,t_{0}))^{n-1-l} \nonumber \\ 
&& - P(t)\sum_{l=0}^{n}(-1)^l \{ \tilde{Q}(t)(\tilde{{\cal C}}(t)-1) \}^l \tilde{Q}(t)(\tilde{{\cal D}}(t)-1)
     (\tilde{Q}(t)\Sigma(t,t_{0}))^{n-l} \nonumber \\
&& - P(t)\sum_{l=0}^{n}(-1)^l \{ (\tilde{{\cal C}}(t)-1)\tilde{Q}(t) \}^{l+1} (\tilde{{\cal D}}(t)-1)
     (\tilde{Q}(t)\Sigma(t,t_{0}))^{n-l},
\end{eqnarray}
where
\begin{eqnarray}
\tilde{Q}(t) &=& U^{-1}_{0}(t,t_{0})Q(t_{0})U^{Q}_{0}(t,t_{0}), \\
\tilde{{\cal C}}(t) &=& U^{-1}_{0}(t,t_{0}){\cal C}(t,t_{0})U_{0}(t,t_{0}), \\
\tilde{{\cal D}}(t) &=& (U^{Q}_{0})^{-1}(t,t_{0}){\cal D}(t,t_{0})U^{Q}_{0}(t,t_{0}).
\end{eqnarray}
Here, $n$ is an integer and $n \ge 1$.
In this derivation, we have used the relation
\begin{eqnarray}
Q(t)\tilde{Q}(t) = Q(t),
\end{eqnarray}
which can be proved from the condition (\ref{eqn:QL0P}).
The second and third terms in $P(t)\Sigma(t,t_{0})(Q(t)\Sigma(t,t_{0}))^{n}$ 
and the fourth and fifth terms in 
$P(t)\Sigma(t,t_{0})(Q(t)\Sigma(t,t_{0}))^{n-1}$
cancel.
The fourth and fifth terms in $P(t)\Sigma(t,t_{0})(Q(t)\Sigma(t,t_{0}))^{n}$ 
and the second and third terms in 
$P(t)\Sigma(t,t_{0})(Q(t)\Sigma(t,t_{0}))^{n+1}$
also cancel.
Therefore, only the first term survives.
As a result, all the terms including $\tilde{{\cal D}}(t)$ disappear.

Note that the relation $\Sigma(t,t_{0})=\Sigma(t,t_{0})Q(t)$ can be 
derived from Eq.~(\ref{eqn:QQQ}).
We thus find 
\begin{eqnarray}
\lefteqn{P(t)\Sigma(t,t_{0})\frac{1}{1-\Sigma(t,t_{0})} } && \nonumber \\
&=&  P(t)\Sigma(t,t_{0})\frac{1}{1-Q(t)\Sigma(t,t_{0})} \nonumber \\
&=&  P(t)\Sigma(t,t_{0})\sum_{n=0}^{\infty}(Q(t)\Sigma(t,t_{0}))^n \nonumber \\
&=&  -P(t)\sum_{n=0}^{\infty}[ \{ -\tilde{Q}(t)(\tilde{{\cal C}}(t)-1) \}^n \tilde{Q}(t)
     -\{ -(\tilde{{\cal C}}(t,t_{0})-1)\tilde{Q}(t) \}^{n+1}  ] \nonumber \\
&=&  -P(t)U^{-1}_{0}(t,t_{0})\sum_{n=0}^{\infty}[ \{ -Q(t)({\cal
C}(t,t_{0})-1) \}^n Q(t) \nonumber \\
&&     -\{ -({\cal C}(t,t_{0})-1)Q(t) \}^{n+1}  ]U_{0}(t,t_{0}) \nonumber \\
&=&  -P(t)U^{-1}_{0}(t,t_{0})Q(t)\frac{1}{1+({\cal
C}(t,t_{0})-1)Q(t)}U_{0}(t,t_{0}) \nonumber \\
&&     -P(t)U^{-1}_{0}(t,t_{0})({\cal C}(t,t_{0})-1)Q(t)\frac{1}{1+({\cal C}(t,t_{0})-1)Q(t)}U_{0}(t,t_{0}) \nonumber \\
&=&  -P(t)U^{-1}_{0}(t,t_{0}){\cal C}(t,t_{0})Q(t)\frac{1}{1+({\cal
C}(t,t_{0})-1)Q(t)}U_{0}(t,t_{0}). \nonumber \\
\end{eqnarray}

\section{Calculation of Correlation Functions Based on Nonequilibrium
TFD}\label{app:5}

To calculate the correlation functions (\ref{eqn:rhop}), 
(\ref{eqn:rhop2}) and (\ref{eqn:rhop3}), 
it is convenient to use a technique of thermo-field dynamics (TFD).
In TFD, the statistical average is expressed as a kind of a vacuum 
expectation value.
As an example, we consider the statistical average of the system with 
the free Hamiltonian of a boson system, 
$H = \sum \omega_{\bf k}d^{\dagger}_{\bf k}d_{\bf k}$.
First, we introduce the transformation through which 
the new pairs of creation and annihilation operators 
$D_{\bf k}$ and $D^{\dagger}_{\bf k}$, 
and $\tilde{D}_{\bf k}$ and $\tilde{D}^{\dagger}_{\bf k}$ are defined:
\begin{eqnarray}
d^{\dagger}_{\bf k}
&=&       
{\rm cosh}\theta_{\bf k}D^{\dagger}_{\bf k} 
+{\rm sinh}\theta_{\bf k}\tilde{D}_{\bf k}, \\
d_{\bf k} 
&=& 
{\rm cosh}\theta_{\bf k}D_{\bf k}
+{\rm sinh}\theta_{\bf k}\tilde{D}^{\dagger}_{\bf k},
\end{eqnarray}
where
\begin{eqnarray}
{\rm sinh}^2 \theta_{\bf k} = n_{\bf k},~~
{\rm cosh}^2 \theta_{\bf k} = 1 + n_{\bf k}. 
\end{eqnarray}
Here, $n_{\bf k}$ is a Bose distribution function.
The above operators satisfy the commutation relation
\begin{eqnarray}
[D_{\bf k}, D^{\dagger}_{{\bf k}'}] &=& 
{\protect [\tilde{D}_{\bf k}, \tilde{D}^{\dagger}_{{\bf k}'}]} 
= \delta^{(3)}_{\bf k,k'}.
\end{eqnarray}
All other commutators vanish.
Now, we can define the thermal vacuum $| \theta_{D} \rangle$ as 
\begin{eqnarray}
D_{\bf k}| \theta_{D} \rangle = \tilde{D}_{\bf k}| \theta_{D} \rangle = 0.
\end{eqnarray}
We thus express the statistical average in terms of the vacuum expectation
value: 
\begin{eqnarray}
{\rm Tr}[\rho_{th} O] &=& \langle \theta_{D} | O | \theta_{D} \rangle, 
\end{eqnarray}
where
\begin{eqnarray}
\rho_{th} &=& e^{-\beta H}/Z, \\
Z &=& {\rm Tr}~e^{-\beta H}.
\end{eqnarray}

Next, we return to the calculation of the correlation functions with 
the local equilibrium density matrix.
In the case of a time-dependent temperature, 
we introduce a time-dependent transformation to 
define the new pairs of creation and annihilation operators 
$B_{\bf k}(t)$ and $B^{\dagger}_{\bf k}(t)$, 
and $\tilde{B}_{\bf k}(t)$ and $\tilde{B}^{\dagger}_{\bf k}(t)$: 
\begin{eqnarray}
b^{\dagger}_{\bf k}(t_{0}) &=& {\rm cosh}\theta^{\pi}_{\bf k}(t)
B^{\dagger}_{\bf k}(t) 
+ {\rm sinh}\theta^{\pi}_{\bf k}(t)\tilde{B}_{\bf k}(t), \\
b_{\bf k}(t_{0}) &=& {\rm cosh}\theta^{\pi}_{\bf k}(t)
B_{\bf k}(t) + {\rm sinh}\theta^{\pi}_{\bf k}(t)\tilde{B}^{\dagger}_{\bf k}(t),
\end{eqnarray}
where
\begin{eqnarray}
{\rm sinh}^2 \theta^{\pi}_{\bf k}(t) 
&=& n^{\pi}_{\bf k}(t), \\
{\rm cosh}^2 \theta^{\pi}_{\bf k}(t) &=& 1 + n^{\pi}_{\bf k}(t).
\end{eqnarray}
Here, $n^{\pi}_{\bf k}(t)$ is the time-dependent distribution function 
of the $\pi$ boson:
\begin{eqnarray}
n^{\pi}_{\bf k}(t) &=& {\rm Tr}[\rho_{E}(t)b^{\dagger}_{\bf
k}(t_{0})b_{\bf k}(t_{0})] \nonumber \\
&=& \frac{1}{e^{\beta(t)\omega^{\pi}_{\bf k}}-1}.
\end{eqnarray}
The creation and annihilation operators 
$B_{\bf k}(t)$, $B^{\dagger}_{\bf k}(t)$, 
$\tilde{B}_{\bf k}(t)$ and $\tilde{B}^{\dagger}_{\bf k}(t)$ 
satisfy 
\begin{eqnarray}
[B_{\bf k}(t),B^{\dagger}_{\bf k'}(t)]
=
[\tilde{B}_{\bf k}(t),\tilde{B}^{\dagger}_{\bf k'}(t)] 
= \delta^{(3)}_{\bf k,k'}.
\end{eqnarray}
All other commutators vanish.
Then, we can define the time-dependent vacuum $|\theta_{B}(t)\rangle$ 
as
\begin{eqnarray}
B_{\bf k}(t)|\theta_{B}(t)\rangle 
= \tilde{B}_{\bf k}(t)|\theta_{B}(t)\rangle 
=0.
\end{eqnarray}
We thus express the statistical average with the local equilibrium 
density matrix in terms of the time-dependent vacuum expectation value:
\begin{eqnarray}
{\rm Tr}[\rho_{E}(t)O] = \langle \theta_{B}(t)|O|\theta_{B}(t)\rangle,
\end{eqnarray}
where $\rho_{E}(t)$ is defined in Eq.~(\ref{eqn:led}).
Therefore, we can simplify the calculation of the correlation 
functions with the help of Wick's theorem.\cite{ref:Gross}
Simply quoting the result, we obtain 
\begin{eqnarray}
{\rm Tr}[\rho_{E}(t) \rho_{\bf k}(t,t_{0})]
&=& \langle \theta_{B}(t)| \rho_{\bf k}(t,t_{0})  |\theta_{B}(t)
\rangle \nonumber \\
&=& \int d^3 {\bf x}e^{i{\bf kx}}
\ccontra{\pi({\bf x},t)}{\pi({\bf x},t)}, \\
{\rm Tr}[\rho_{E}(t) [\rho_{\bf k_{1}}(s,t_{0}),\rho_{\bf k_{2}}(t,t_{0})]]
&=& \langle \theta_{B}(t) |[\rho_{\bf l}(s,t_{0}),\rho_{\bf
k}(t,t_{0})] |\theta_{B}(t) \rangle \nonumber \\
&=& \int d^3 {\bf x}_{1}d^3 {\bf x}_{2}
e^{i{\bf k}_{1}{\bf x}_{1}}e^{i{\bf k}_{2}{\bf x}_{2}} \nonumber \\
&& \times 2 \{
(\ccontrad{\pi({\bf x}_{1},s)}{\pi({\bf x}_{2},t)}
\hspace{1.2cm}
)^2
-(\ccontrad{\pi({\bf x}_{2},t)}{\pi({\bf x}_{1},s)}
\hspace{1.3cm}
)^2
\},\nonumber \\
\end{eqnarray}
where
\begin{eqnarray}
\ccontradd{\pi({\bf x}_{1},t_{1})}{\pi({\bf x}_{2},t_{2})}
\hspace{1.3cm}
&=& \sum_{\bf k_{1},k_{2}}\frac{1}{2V\sqrt{\omega^{\pi}_{\bf
k_{1}}\omega^{\pi}_{\bf k_{2}} }}[\cosh^{\pi}_{\bf k_{1}}(t) B_{\bf k_{1}}(t)
e^{-i\omega^{\pi}_{\bf k_{1}}(t_{1}-t_{0}) }+ \sinh
\theta^{\pi}_{\bf -k_{1}}(t)\tilde{B}_{\bf -k_{1}}(t)
e^{i\omega^{\pi}_{\bf k_{1}}(t_{1}-t_{0}) }, \nonumber \\
&& \sinh \theta^{\pi}_{\bf -k_{2}}(t)\tilde{B}^{\dagger}_{\bf -k_{2}}(t)
e^{-i\omega^{\pi}_{\bf k_{2}}(t_{2}-t_{0}) } 
+ \cosh^{\pi}_{\bf k_{2}}(t) B^{\dagger}_{\bf k_{2}}(t)
e^{i\omega^{\pi}_{\bf k_{2}}(t_{2}-t_{0}) }] \nonumber \\
&=& \sum_{\bf k}\frac{1}{2V\omega^{\pi}_{\bf k}}
(
\cosh^2 \theta^{\pi}_{\bf k}(t) e^{-i\omega^{\pi}_{\bf k}(t_{1}-t_{2})}
+\sinh^2 \theta^{\pi}_{\bf k}(t) e^{i\omega^{\pi}_{\bf k}(t_{1}-t_{2})}
)e^{i{\bf k}({\bf x}_{1}-{\bf x}_{2})}.
\end{eqnarray}
Readers familiar with nonequilibrium thermo-field dynamics 
may notice the difference between the technique that we have outlined 
here and that used in Ref.~\cite{ref:NETFD}.
However, these two techniques give the same result with respect to 
the trace calculation.\cite{PC:Yama}

\end{document}